\documentclass{article}

\usepackage{microtype}
\usepackage{graphicx}
\usepackage{subfigure}
\usepackage{graphicx}
\usepackage{amsmath}
\usepackage{amssymb}
\usepackage{authblk}
\usepackage{hyperref}
\usepackage{cleveref}
\usepackage{lmodern}
\usepackage{enumitem}
\usepackage{booktabs}
\usepackage{wrapfig}

\usepackage{hyperref}

\newcommand{\sysname}{PUD-LLM}

\usepackage[accepted]{mlsys2025}

\makeatletter
\renewcommand{\Notice@String}{Preprint.}
\makeatother

\mlsystitlerunning{Accelerating LLM Decoding with In-DRAM Parallelism}

\begin{document}

\twocolumn[
% \mlsystitle{\sysname: Unlocking Parallelism in Processing-using-DRAM through Layer Redundancy \texorpdfstring{\\}{ } for Fast LLM Decoding}
\mlsystitle{Accelerating LLM Decoding with In-DRAM Parallelism}

% \mlsystitle{\sysname: Optimizating on Processing-Using-DRAM}

% It is OKAY to include author information, even for blind
% submissions: the style file will automatically remove it for you
% unless you've provided the [accepted] option to the mlsys2025
% package.

% List of affiliations: The first argument should be a (short)
% identifier you will use later to specify author affiliations
% Academic affiliations should list Department, University, City, Region, Country
% Industry affiliations should list Company, City, Region, Country

% You can specify symbols, otherwise they are numbered in order.
% Ideally, you should not use this facility. Affiliations will be numbered
% in order of appearance and this is the preferred way.
\mlsyssetsymbol{equal}{*}

\begin{mlsysauthorlist}
\mlsysauthor{Kevin Xu}{ut}
\mlsysauthor{Tatsuya Kubo}{ut}
\mlsysauthor{Daichi Tokuda}{ut}
\mlsysauthor{Shinya Takamaeda-Yamazaki}{ut,riken}
\mlsysauthor{Issei Sato}{ut}
\end{mlsysauthorlist}

\mlsysaffiliation{ut}{Department of Computer Science, Graduate School of Information Science and Technology, The University of Tokyo, Tokyo, Japan}

\mlsysaffiliation{riken}{RIKEN Center for Advanced
Intelligence Project (AIP), Japan}

\mlsyscorrespondingauthor{Kevin Xu}{kevinxu@g.ecc.u-tokyo.ac.jp}
%\mlsyscorrespondingauthor{Tatsuya Kubo}{tatsuya.kubo@is.s.u-tokyo.ac.jp}
%\mlsyscorrespondingauthor{Daichi Tokuda}{tokuda-daichi@is.s.u-tokyo.ac.jpp}
%\mlsyscorrespondingauthor{Shinya Takamaeda-Yamazaki}{shinya@is.s.u-tokyo.ac.jp}
%\mlsyscorrespondingauthor{Issei Sato}{sato@g.ecc.u-tokyo.ac.jp}

% You may provide any keywords that you
% find helpful for describing your paper; these are used to populate
% the "keywords" metadata in the PDF but will not be shown in the document
\mlsyskeywords{Machine Learning, MLSys}

\vskip 0.3in

\begin{abstract}
Autoregressive decoding of large language models (LLMs) on edge devices is dominated by memory-bound general matrix--vector multiplication (GeMV) operations. Processing-using-DRAM (PUD) is a promising computing paradigm that can alleviate this bottleneck by reducing processor--memory data transfers and enabling massively parallel computation within DRAM (e.g., across 65,536 columns). However, weight matrices in individual LLM layers (e.g., a QKV matrix in Llama~3~8B with 6,144 rows) are typically too small to fully utilize parallelism in PUD. To address this challenge, we propose \sysname{}, an algorithm--system co-design for PUD-based LLM inference. At the algorithm level, we enable a subsequent projection to be packed with the current projection by reformulating it as a computation on the current residual state plus a correction for the residual update. Layer redundancy allows this correction to be approximated using fewer bit-serial steps. At the system level, we tailor data mapping and execution to improve utilization of either columns within DRAM subarrays (SIMD) or independently controlled banks (MIMD). Overall, \sysname{} achieves up to $1.41\times$ end-to-end decoding speedup over the PUD baseline, corresponding to a $2.45\times$ speedup over processor-based execution.
\end{abstract}
]
%At the algorithm level, we decompose a subsequent projection into a lookahead computation and a delta correction. The lookahead are packed alognside currnet projection. The lookahead overlaps with the current projection on otherwise idle PUD resources, while layer redundancy allows the delta correction to be approximated using fewer bit-serial steps.

\printAffiliationsAndNotice{}  % leave blank if no need to mention equal contribution
% \printAffiliationsAndNotice{\mlsysEqualContribution} % otherwise use the standard text.

%%%%%%%%%%%%%%%%%%%%%%%%%%%%%%%%%%%%%%%%%%%%%%%%%%%%%
\section{Introduction}
\label{sec:introduction}

Efficient inference for large language models (LLMs) is important for edge devices with limited power and hardware resources. Autoregressive decoding on these devices runs at batch size one, where each linear projection becomes a matrix--vector multiplication (GeMV) between a weight matrix and an activation vector. These operations have low arithmetic intensity because they perform little computation per weight fetched from memory. Consequently, decoding on CPUs and GPUs is typically bottlenecked by weight transfers, leaving much of their computational capacity unused~\cite{shazeer2019fast,pope2023efficiently,gholami2024ai}. In this regime, increasing computational throughput alone offers limited improvements in decoding latency.

Processing-in-memory (PIM) is a promising computing paradigm to address this bottleneck by performing computation near or within memory, reducing costly data movement~\cite{mutlu2025memory}. Processing-near-memory places dedicated compute units close to memory arrays~\cite{devaux2019true}. However, integrating compute units into DRAM incurs significant area overhead and increases design complexity under the constraints of DRAM fabrication. In contrast, processing-using-DRAM (PUD) is a promising computing paradigm that leverages the operational principles of existing DRAM cells and sense amplifiers to perform massively parallel computation directly within DRAM arrays. % Notably, PUD can be implemented with minimal DRAM chip area overhead~\cite{seshadri2017ambit}, or even on unmodified commodity DRAM~\cite{gao2019computedram}. 

\begin{figure}[t]
    \centering
    \includegraphics[width=0.95\linewidth]{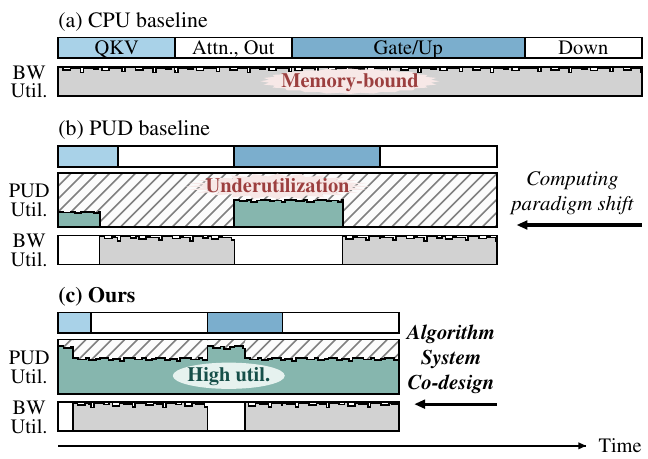}
   \caption{PUD is a promising computing paradigm for memory-bound LLM decoding. \sysname{} unlocks its underutilized parallelism through algorithm--system co-design.}
\end{figure}

PUD exploits DRAM's column-level parallelism to achieve high computational throughput~\cite{seshadri2013rowclone,seshadri2017ambit}. DRAM command sequences implement bitwise primitives that are combined into bit-serial arithmetic operations~\cite{seshadri2017ambit,gao2019computedram}. Each primitive operates across many DRAM columns simultaneously, enabling single-instruction, multiple-data (SIMD) execution~\cite{hajinazar2021simdram}. Independently controlled regions, such as channels and banks, further provide multiple-instruction, multiple-data (MIMD) parallelism~\cite{kim2012case,oliveira2024mimdram}. Although bit-serial arithmetic requires long sequences of DRAM commands, prior work~\cite{kubo2025mvdram,de2026count2multiply} demonstrates that PUD can outperform CPU and GPU execution for sufficiently large low-bit matrix--vector multiplications by leveraging its massive parallelism and exploiting bit sparsity to reduce the number of required DRAM commands.

However, LLM matrix sizes are typically insufficient to fully utilize this parallelism. Moreover, the sequential dependencies between layers prevent overlapping their execution during batch-size-one decoding. As a result, PUD parallelism remains underutilized, significantly limiting end-to-end LLM inference performance. For example, the QKV projection in Llama~3.1~8B~\cite{grattafiori2024llama}, with a combined output dimension of 6,144, uses only 24,576 of the 65,536 available DRAM columns with 4-bit weights, leaving 62.5\% of the column-level parallelism unused.

System-level optimizations alone cannot fully resolve this challenge while preserving the original computational dependencies. Prior work improves PIM utilization through data placement~\cite{ibrahim2024pimnast}, fine-grained resource control~\cite{oliveira2024mimdram}, and scheduling across independent sub-batches~\cite{heo2024neupims}. However, these approaches preserve the sequential dependencies between consecutive LLM sublayers. Consequently, parallel computations remain limited to operations within the current layer, leaving substantial PUD resources underutilized.

\begin{figure}[t]
    \centering
    \includegraphics[width=0.95\linewidth]{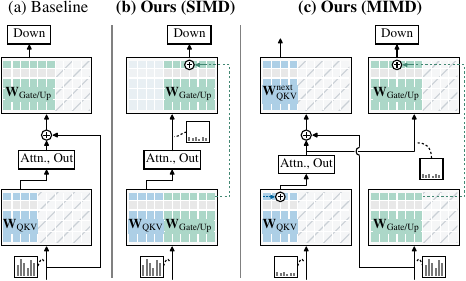}
    \caption{Our system design improves column (SIMD) utilization through packing and bank (MIMD) utilization through pipelining.}
    \label{fig:lookahead_mechanisms}
\end{figure}

To address this challenge, we propose \sysname{}, a hardware-aware algorithm and system co-design for PUD-based LLM inference. At the algorithm level, we use linearity to decompose a subsequent projection into a lookahead computation on the current residual state and a correction based on the residual update. Although this decomposition increases the total computational work, it reduces decoding latency in two ways. First, the lookahead computation overlaps with the preceding sublayer by exploiting otherwise idle PUD resources. Second, the similarity between consecutive representations in LLMs, known as layer redundancy~\cite{men2024shortgpt}, allows the correction to be approximated using fewer bit-serial computation steps.

At the system level, we develop two execution schemes to improve utilization of PUD's column-level (SIMD) and bank-level (MIMD) parallelism as illustrated in Figure~\ref{fig:lookahead_mechanisms}. The SIMD scheme reformulates Gate/Up lookahead to share QKV's input and maps its weights to otherwise unused DRAM columns, allowing both projections to execute under a common instruction stream. The MIMD scheme applies the decomposition to both QKV and Gate/Up, allowing each projection's correction and its successor's lookahead to execute in parallel on independently controlled banks or channels. These schemes improve utilization within subarrays and across independent resources, respectively.

In summary, our contributions are as follows:
\begin{itemize}[leftmargin=*]
    \item We identify how individual projection dimensions and sequential sublayer dependencies leave PUD parallelism underutilized in PUD-based LLM decoding.
    \item We propose a training-free algorithm that exposes parallelism across sublayers and exploits inter-layer redundancy to reduce bit-serial steps on the critical path.
    \item We develop system-level optimizations for two execution schemes that exploit unused SIMD or MIMD parallelism available in a given model--memory configuration.
    \item \sysname{} achieves up to $1.41\times$ end-to-end decoding speedup over the PUD baseline, with $1.85\times$ higher mean PUD resource utilization.
\end{itemize}

%%%%%%%%%%%%%%%%%%%%%%%%%%%%%%%%%%%%%%%%%%%%%%%%%%%%%
\section{Background and Motivation}
\label{sec:background}

This section reviews Transformer-based LLMs, DRAM, and PUD, highlighting how PUD executes GeMV and why its parallelism remains underutilized during LLM decoding.

%%%%%%%%%%%%%%%%%%%%%%%%%%%
\subsection{Transformer-based LLM}
\label{subsec:transformer}

%\begin{figure}[ht]
%  \centering
%  \includegraphics[width=\linewidth]{figures/transformer_block.pdf}
%  \vspace{-5.5mm}
%  \caption{Transformer model.}
%\end{figure}

A Transformer model~\cite{vaswani2017attention} consists of a stack of blocks, each containing an attention sublayer followed by a feed-forward network (FFN) sublayer. Both sublayers share a common structure: input linear projections, intermediate nonlinear processing, and an output linear projection. In attention, the input is projected into queries, keys, and values, followed by self-attention and an output projection back to the model's hidden dimension. A common FFN design uses gate and up projections, followed by nonlinear gating, element-wise multiplication, and a down projection back to the hidden dimension~\cite{shazeer2020glu}. Each sublayer operates on a normalized input and adds its output back to the original input via a residual connection:
\begin{equation}
  \mathbf{x}'
  =
  \mathbf{x}
  +
  \mathbf{W}_{\mathrm{out}}
  \Phi\!\left(
    \mathbf{W}_{\mathrm{in}}
    \mathrm{Norm}(\mathbf{x})
  \right),
  \label{eq:bg_residual_form}
\end{equation}
where $\mathbf{x}$ and $\mathbf{x}'$ denote the residual states before and after the update, respectively. The matrix $\mathbf{W}_{\mathrm{in}}$ vertically concatenates the weight matrices of the input projections, $\Phi$ denotes the intermediate processing such as attention or nonlinear gating, and $\mathbf{W}_{\mathrm{out}}$ denotes the output projection. This expression summarizes a standard sublayer; architectural variants may incorporate expert routing~\cite{jiang2024mixtral} or alternative attention~\cite{team2026qwen3}. %,deepseek2025v32}.

%%%%%%%%%%%%%%%%%%%%%%%%%%%
\subsection{Processing-using-DRAM}
\label{subsec:pud}

\begin{figure}[t]
    \centering
    \includegraphics[width=0.95\linewidth]{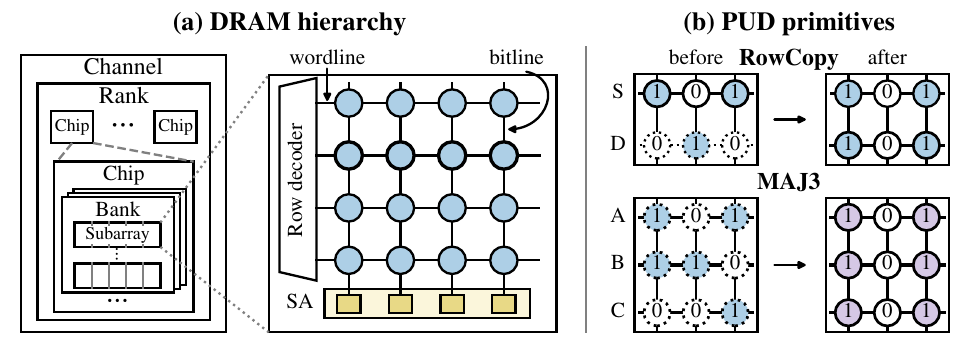}
    \caption{(a) DRAM organization and (b) PUD primitives.} % Column-parallel row copying (RowCopy) and three-input majority (MAJ3) operations.}
    \label{fig:dram}
\end{figure}
\paragraph{DRAM Organization and Operation.} A DRAM memory system comprises channels, ranks, chips, banks, and subarrays, as shown in Figure~\ref{fig:dram}(a)~\cite{jacob2010memory}. Each rank groups chips that operate in lockstep under shared command and address signals. Within a subarray, one-bit cells are arranged in rows and columns. Cells in each row share a wordline, while those in each column connect to a bitline and its sense amplifier (SA). The SAs collectively form the local row buffer. An ACT command opens a row, allowing the SAs to sense, amplify, and restore its data. READ and WRITE commands transfer selected portions of the open row over the data bus, while PRE closes the row and precharges the bitlines for the next activation.

\paragraph{In-DRAM Computation.} PUD exploits the analog behavior of DRAM cells and sense amplifiers to perform row copying and bitwise logic without transferring data outside DRAM (Figure~\ref{fig:dram}(b)). Row copying transfers data between rows within the same subarray through shared sense amplifiers and bitlines~\cite{seshadri2013rowclone}, and simultaneous activation of multiple rows enables majority operations~\cite{seshadri2017ambit}. These operations can be implemented using timing-violating command sequences in unmodified DRAM (u-PUD)~\cite{gao2019computedram} or additional hardware support in modified DRAM (m-PUD)~\cite{seshadri2017ambit}. Multi-bit arithmetic is implemented through sequences of these bitwise operations. Applying a common operation across many columns provides single-instruction, multiple-data (SIMD) parallelism~\cite{hajinazar2021simdram}. Interleaving commands to different banks over a channel's shared command/address bus allows independent operation sequences to execute concurrently, providing multiple-instruction, multiple-data (MIMD) parallelism~\cite{kim2012case,oliveira2024mimdram}.
%
%\begin{figure}[t]
%    \centering
%    \includegraphics[width=\linewidth, height=3.5cm]{example-image-a}
%    \vspace{-5.5mm}
%    \caption{SIMD and MIMD paralllism hogge.}
%    \label{fig:dram}
%\end{figure}

%%%%%%%%%%%%%%%%%%%%%%%%%%%
\subsection{GeMV Execution in DRAM}

\begin{figure}[t]
    \centering
    \includegraphics[width=0.95\linewidth]{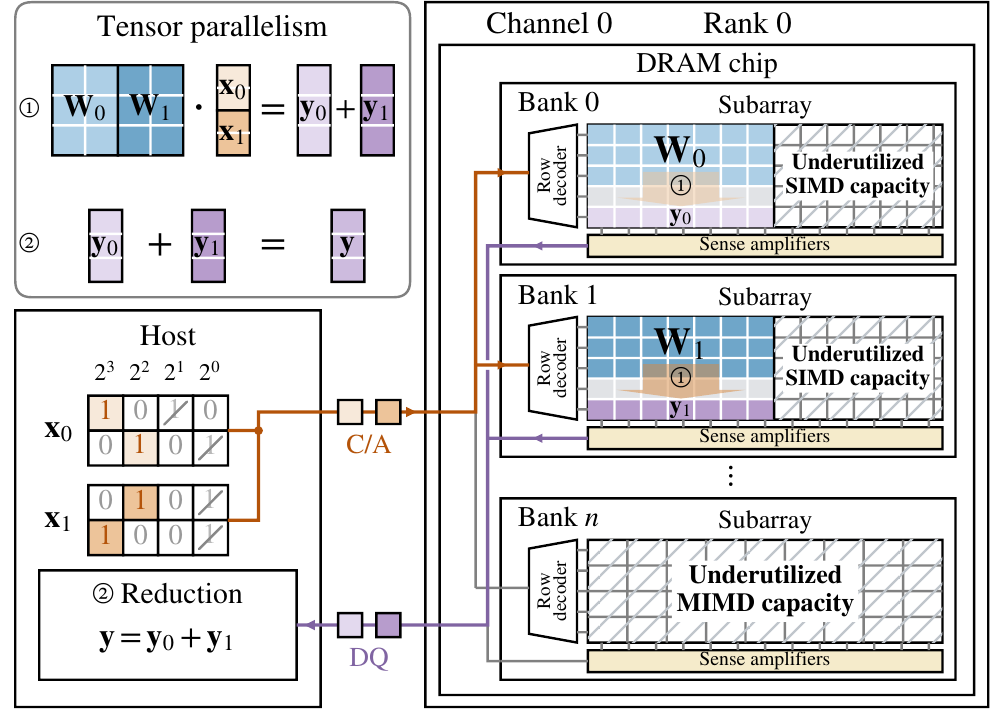}
    \caption{PUD-based GeMV consists of two stages: (1) in-DRAM computation with cost proportional to the number of retained nonzero input terms; (2) partial-sum readout for reduction.}
    \label{fig:gemv_pud}
\end{figure}

\paragraph{Execution Principle.} Prior work~\cite{kubo2025mvdram,de2026count2multiply} demonstrates that PUD can outperform processor-based execution for sufficiently large low-bit matrix--vector multiplications by exploiting massive parallelism and input bit-level sparsity. We follow the data mapping and execution scheme of MVDRAM~\cite{kubo2025mvdram}, as illustrated in Figure~\ref{fig:gemv_pud}. Consider $\mathbf{y}=\mathbf{W}\mathbf{x}$, where $\mathbf{W}\in\mathbb{R}^{M\times K}$ has input dimension $K$ and output dimension $M$. 
Each weight vector $\mathbf{W}_{:,k}$ is mapped to a DRAM row, with output dimensions and weight-bit positions mapped to columns. For $b_w$-bit weights, each row requires $Mb_w$ columns. Thus, output dimension and weight precision determine SIMD utilization for a given column capacity.
Execution consists of PUD computation and host-side reduction. During computation, the host translates each nonzero bit of $x_k$ into DRAM commands that accumulate the appropriately scaled contribution of $\mathbf{W}_{:,k}$ from its stored row, exploiting input bit-level sparsity. During reduction, the partial sums within DRAM are transferred to the host and combined to obtain the output.

\begin{figure}[t]
    \centering
    \includegraphics[width=0.95\linewidth]{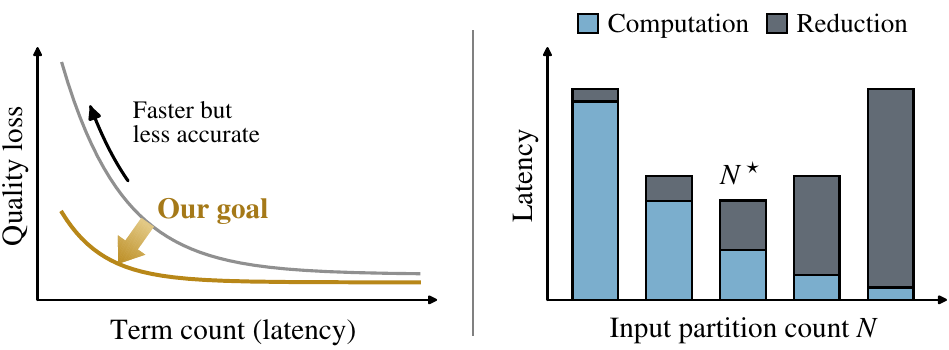}
    \caption{Left: Schematic quality--efficiency trade-off in bit-serial computation. Right: Partition-dependent latency of PUD.}
    \label{fig:background_tradeoffs}
\end{figure}

\paragraph{Quality--Efficiency Trade-off.} In this PUD GeMV execution scheme, computation time scales approximately with the number of nonzero terms. Using more terms improves numerical accuracy but increases latency, creating a quality--efficiency trade-off, as illustrated in Figure~\ref{fig:background_tradeoffs} (left). This trade-off can be controlled at runtime by truncating less significant terms under a specified budget~\cite{kung2020term}. Our goal is to improve this Pareto frontier by optimizing end-to-end LLM inference beyond individual GeMV kernels.

\paragraph{Tensor Parallelism.} The input dimensions of LLM weight matrices exceed the row capacity of a DRAM subarray (typically 512--1024 rows), requiring input partitioning. Input partitioning is a form of tensor parallelism~\cite{shoeybi2019megatron}, decomposing the projection into independent partial products: $\mathbf{y}=\sum_{n=1}^{N}\mathbf{W}_n\mathbf{x}_n$. These partial products can execute concurrently on independently controlled banks or channels, exploiting MIMD parallelism. Each partition processes $K/N$ input elements, where $N$ is the input partition count. Increasing $N$ therefore reduces computation per partition but increases the number of partial sums to read and aggregate, creating a trade-off between computation and reduction.

\paragraph{Partitioning Optimization.}
We model this trade-off to select the input partition count that minimizes GeMV latency. Let $\tau$ denote the average number of retained nonzero terms per input element. Assuming $N$ is large enough to saturate the readout bandwidth but small enough for all partitions to execute concurrently, latency is approximated by
\begin{equation}
    T(N)=\frac{\alpha K\tau}{N}+\beta MN,
    \label{eq:bg_partition_latency}
\end{equation}
where $\alpha$ is the PUD computation cost per term and $\beta$ is the reduction cost per output element per partition. Figure~\ref{fig:background_tradeoffs} (right) illustrates this trade-off. The unconstrained continuous optimum is $N^\star=\sqrt{\alpha K\tau/(\beta M)}$, at which computation and reduction contribute equally. The cost coefficients $\alpha$ and $\beta$ are detailed in Appendix~\ref{app:pud_execution}. The selected partition count determines how many independent resources can be used, and hence MIMD utilization.

%%%%%%%%%%%%%%%%%%%%%%%%%%%
\subsection{Challenges in PUD-Based LLM Inference}
\label{sec:pud_underutilization}

\begin{figure}[t]
    \centering
    \includegraphics[width=0.94\linewidth]{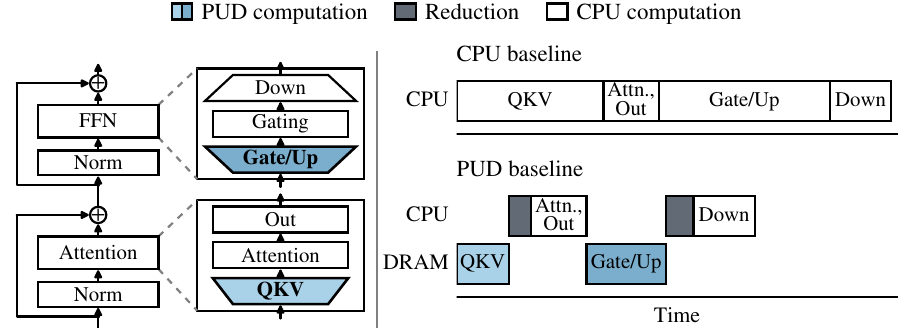}
    \caption{
        PUD mapping and execution of a Transformer block.
    }
    \label{fig:pud_bottleneck}
\end{figure}
Figure~\ref{fig:pud_bottleneck} shows the mapping and execution of a Transformer block on PUD. We offload QKV and Gate/Up to PUD, exploiting their wide output dimensions to accelerate these projections through column parallelism, while the remaining operations execute on the CPU. However, QKV and Gate/Up execute sequentially due to dependencies between sublayers. Each projection's output dimension and weight precision determine its column demand, while its input partition count determines the number of execution resources used concurrently. When these demands fall below the available capacity, SIMD or MIMD parallelism remains underutilized. We illustrate these two cases using Llama~3.1~8B on two memory configurations.

\paragraph{Case 1: Underutilized SIMD Parallelism.} On a u-PUD system with one 8\,GiB single-rank DDR4 DIMM and 16 banks on a single channel, 4-bit Llama~3.1~8B uses 16 latency-optimal input partitions for QKV, occupying all 16 banks. However, its 6,144 output dimensions require only 24,576 of the 65,536 available columns per partition, leaving 62.5\% of the SIMD capacity unused.

\paragraph{Case 2: Underutilized MIMD Parallelism.} On an m-PUD system with one 16\,GiB dual-rank DDR4 DIMM per channel across two channels, the same model's QKV and Gate/Up projections occupy 32 and 30 of the 42 effective MIMD resources, respectively, under latency-optimal input partitioning (Appendix~\ref{app:pud_execution}). This leaves 23.8\% and 28.6\% of the MIMD capacity unused.

%%%%%%%%%%%%%%%%%%%%%%%%%%%%%%%%%%%%%%%%%%%%%%%%%%%%%
\section{Algorithm Design}
\label{sec:algorithm}

This section presents an algorithm that exploits inter-layer redundancy in LLMs to expose parallel work for PUD.

\begin{figure}[t]
    \centering
    \includegraphics[width=0.95\linewidth]{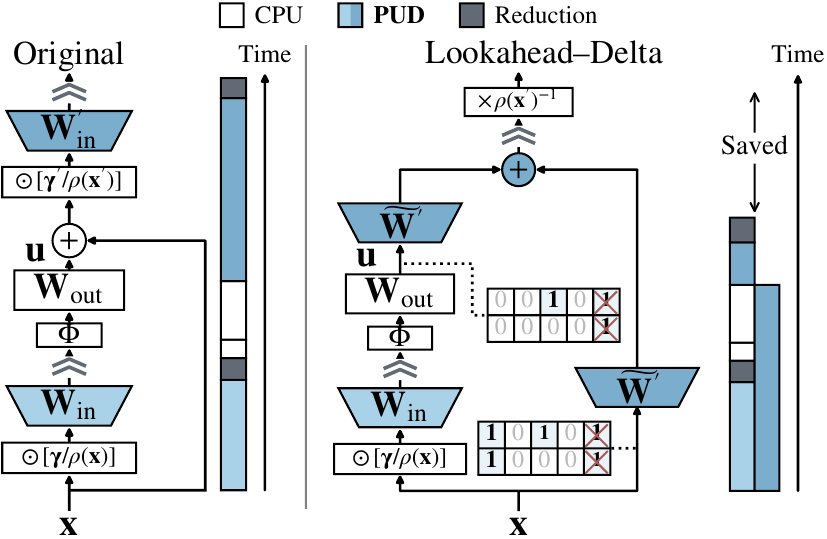}
    \caption{Our algorithm decomposes a subsequent projection into lookahead that uses PUD parallelism to execute alongside the current projection and a correction approximated with fewer terms by exploiting layer redundancy.}
    \label{fig:ld_dataflow}
\end{figure}

\subsection{Algebraic Transformation}
% \subsection{Lookahead--Delta Decomposition}
\label{sec:ld_decomposition}

Let $\mathbf{x}$ denote the current residual state and $\mathbf{u}$ the update produced by the current sublayer. The successor's input is
\begin{equation}
    \mathbf{x}' = \mathbf{x} + \mathbf{u}.
    \label{eq:ld_residual}
\end{equation}
Ordinarily, the successor's input projection of $\mathbf{x}'$ must wait until the current sublayer produces $\mathbf{u}$. We restructure this dependency by separating the contributions of $\mathbf{x}$ and $\mathbf{u}$.

% We include the RMSNorm preceding the input projection in this decomposition. 
Let $\mathbf{W}'_{\mathrm{in}}$ and $\boldsymbol{\gamma}'$ denote the successor's projection weights and learned normalization gains, respectively. Writing
$\mathrm{Norm}(\mathbf{x}')
= \operatorname{diag}(\boldsymbol{\gamma}')\mathbf{x}'/\rho(\mathbf{x}')$,
where $\rho$ is the input-dependent RMS normalization factor, we absorb the gains into
$\widetilde{\mathbf{W}}'
= \mathbf{W}'_{\mathrm{in}}\operatorname{diag}(\boldsymbol{\gamma}')$.
By linearity, we obtain
\begin{equation}
\label{eq:ld_decomposition}
\mathbf{W}'_{\mathrm{in}}\mathrm{Norm}(\mathbf{x}')
=
\frac{1}{\rho(\mathbf{x}')}
\left(
\underbrace{\widetilde{\mathbf{W}}'\mathbf{x}}_{\text{lookahead}}
+
\underbrace{\widetilde{\mathbf{W}}'\mathbf{u}}_{\text{delta}}
\right).
\end{equation}
As illustrated in Figure~\ref{fig:ld_dataflow}, the decomposition enables lookahead to run concurrently with the current sublayer using the already available state $\mathbf{x}$. Although it splits one projection into two computations, it requires only one readout and reduction. To achieve this, we retain the lookahead partial results in DRAM without readout or reduction. Once $\mathbf{u}$ becomes available, we compute delta and accumulate it into these partial results. We then read out and reduce the combined results and apply the normalization factor.

\paragraph{Why Is This Algorithm Particularly Effective on PUD?}
The benefit of this decomposition depends on the ability to overlap the current projection with the successor's lookahead, which uses a different weight matrix. In memory-bound processor-based execution, these computations compete for weight-transfer bandwidth. In contrast, PUD performs computation within DRAM, allowing lookahead to exploit otherwise idle computational resources without additional weight transfers to the host. This distinction reflects two different forms of parallelism. Token-parallel methods~\cite{leviathan2023fast,fu2024break} exploit processor-side parallelism by processing multiple input vectors with the same weight matrix, increasing weight reuse, whereas our design exploits in-memory parallelism by concurrently executing computations on different weight matrices.

\subsection{Efficient Approximation via Layer Redundancy}
\label{sec:delta_approximation}

The decomposition replaces one matrix--vector product with two, increasing arithmetic work. Even if lookahead is fully overlapped, the decomposition alone does not reduce latency. We therefore exploit redundancy across LLM layers to approximate the delta projection with fewer terms.

\begin{figure}[t]
    \centering
    \includegraphics[width=0.95\linewidth]{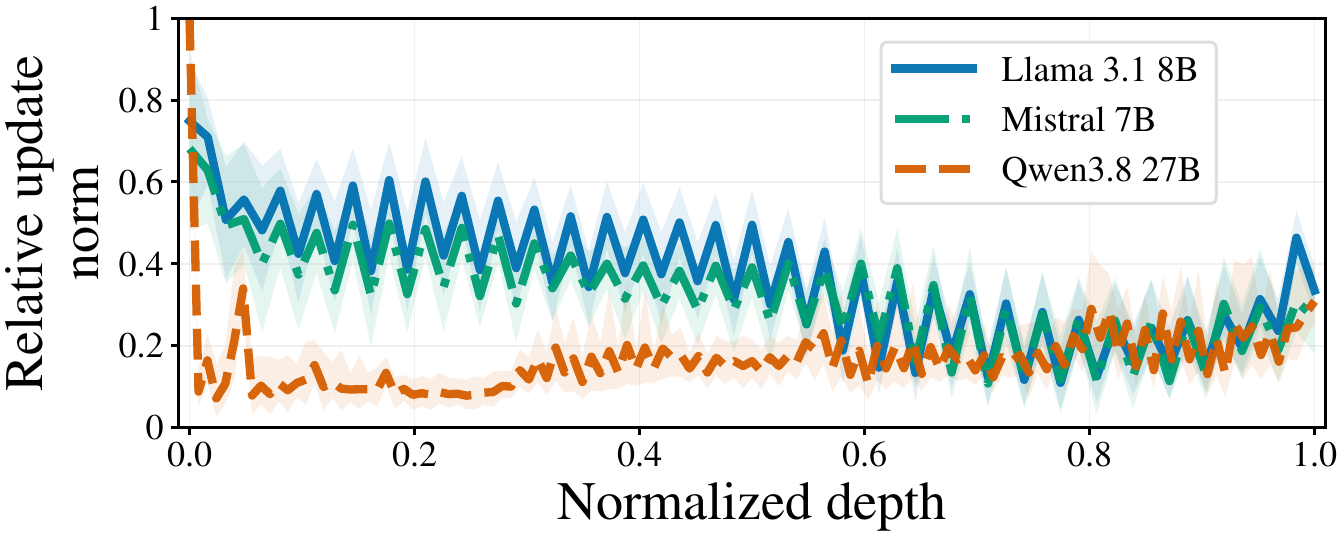}
    \caption{Layer redundancy in LLMs.} 
    \label{fig:layer_redundancy}
\end{figure}

It has been shown that residual states in LLMs change relatively little between consecutive sublayers~\cite{razzhigaev2024secretly,men2024shortgpt}. For a residual update $\mathbf{u}$ and the updated state $\mathbf{x}'=\mathbf{x}+\mathbf{u}$, we define the relative update norm as $\eta=\|\mathbf{u}\|_2/\|\mathbf{x}'\|_2$. Figure~\ref{fig:layer_redundancy} shows how $\eta$ varies across model depth on WikiText-2~\cite{merity2017pointer}.

A small residual update allows a coarser approximation
at the same relative state-error tolerance.
Let $\widehat{\mathbf{v}}_{\tau}$ approximate $\mathbf{v}$
using an average of $\tau$ significant nonzero terms per input element,
with relative error
\(
e_{\mathbf{v}}(\tau)
=
\|\widehat{\mathbf{v}}_{\tau}-\mathbf{v}\|_2
\,/\, \|\mathbf{v}\|_2.
\)
Assuming exact lookahead computation, approximating the update
as $\widehat{\mathbf{u}}_{\tau_{\mathrm{delta}}}$ yields
the reconstructed state
$\widehat{\mathbf{x}}'=\mathbf{x}+\widehat{\mathbf{u}}_{\tau_{\mathrm{delta}}}$
with relative error
\begin{equation}
    \frac{\|\widehat{\mathbf{x}}'-\mathbf{x}'\|_2}
         {\|\mathbf{x}'\|_2}
    =
    \frac{\|\mathbf{u}\|_2}{\|\mathbf{x}'\|_2} \frac{\|\widehat{\mathbf{u}}_{\tau_{\mathrm{delta}}}
             -\mathbf{u}\|_2}
         {\|\mathbf{u}\|_2}
    =
    \eta\,e_{\mathbf{u}}(\tau_{\mathrm{delta}}).
    \label{eq:delta_relative_error}
\end{equation}
Thus, when $\eta<1$, achieving a target relative error $\varepsilon$ in the full state requires an update approximation error of at most $\varepsilon/\eta$, allowing a coarser approximation of the update. To estimate how many fewer terms this permits, we model the relative approximation error as a function of the term count:
\begin{equation}
    e_{\mathbf{v}}(\tau)
    = a_{\mathbf{v}}\exp(-c_{\mathbf{v}}\tau),
    \qquad a_{\mathbf{v}},c_{\mathbf{v}}>0.
    \label{eq:term_error_model}
\end{equation}
For simplicity, we assume common coefficients, i.e., $a_{v} = a$ and $c_{v} = c$. Under this model, matching the relative state error of direct
approximation requires
$\eta\,e_{\mathbf{u}}(\tau_{\mathrm{delta}})
=e_{\mathbf{x}'}(\tau)$, yielding
\begin{equation}
    \tau_{\mathrm{delta}}
    =
    \tau+\frac{\log\eta}{c}.
    \label{eq:delta_term_reduction}
\end{equation}
For $0<\eta<1$, delta approximation requires fewer terms than direct approximation at the same error. When lookahead latency is fully hidden, the reduced cost of delta computation translates into a shorter critical path.

%%%%%%%%%%%%%%%%%%%%%%%%%%%%%%%%%%%%%%%%%%%%%%%%%%%%%
\begin{figure*}[t]
    \centering
    \includegraphics[width=0.95\linewidth]
        {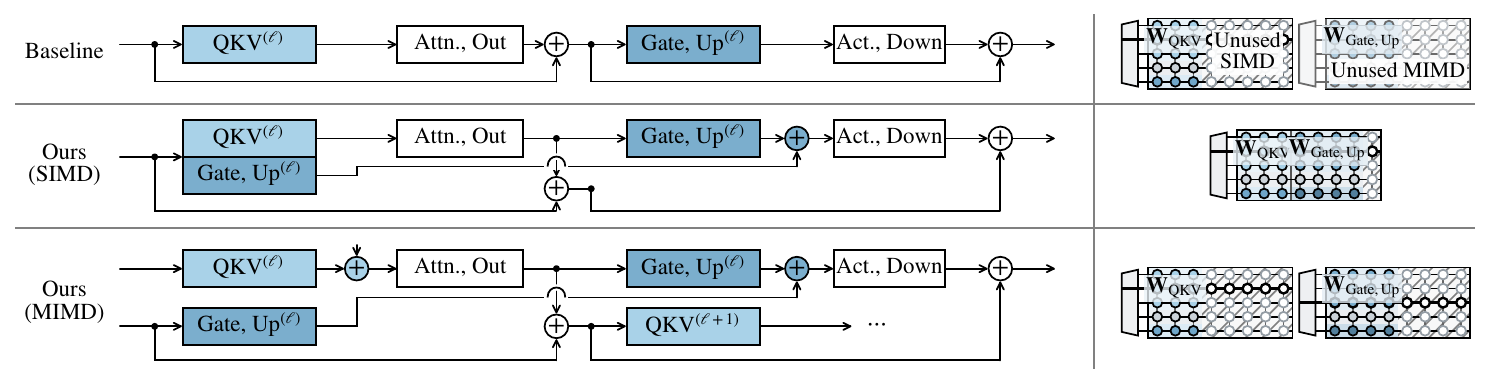}
    \caption{
        Execution flows and data mappings of the PUD baseline and the two proposed schemes. SIMD executes QKV and Gate/Up lookahead under a shared instruction stream. MIMD overlaps the current projection's delta with the successor's lookahead.
    }
    \label{fig:execution_computation_graphs}
\end{figure*}

\section{System Design}
\label{sec:system}

This section describes how to efficiently execute the parallel work exposed by lookahead--delta decomposition. We present two execution schemes tailored to underutilized SIMD and MIMD parallelism, respectively, then describe their data mappings and optimize input partitioning.

%%%%%%%%%%%%%%%%%%%%%%%%%%%
\subsection{Execution Schemes}
\label{sec:execution_schemes}

\begin{figure}[t]
    \centering
    \includegraphics[width=0.95\linewidth]{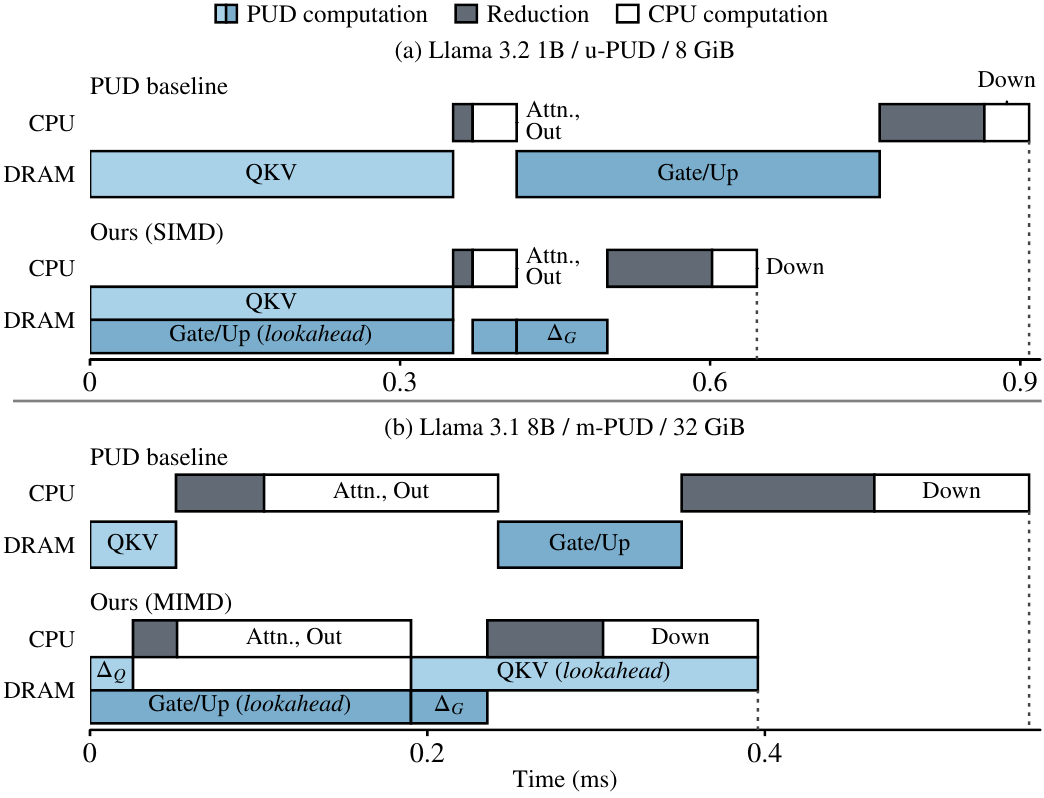}
    \caption{Execution timelines comparing the PUD baseline with (a) SIMD packing and (b) MIMD pipelining.}
    \label{fig:execution_timeline}
\end{figure}

Different model--memory configurations leave different forms of parallelism underutilized. We develop two execution schemes: one uses SIMD parallelism to execute lookahead alongside the current projection, while the other uses MIMD parallelism to pipeline lookahead and delta computations across successive sublayers. Figure~\ref{fig:execution_computation_graphs} shows their execution flows, while Figure~\ref{fig:execution_timeline} illustrates their timelines for representative configurations, using input partitioning optimized as described in the next subsection.

%%%%%%%%%%%%%
\subsubsection{SIMD Execution through Packing}
\label{sec:simd_packing}

We apply lookahead--delta decomposition only to Gate/Up, allowing its lookahead to execute alongside direct QKV computation. SIMD execution requires the two projections to share an input and instruction stream. However, the Gate/Up lookahead in Equation~\eqref{eq:ld_decomposition} uses $\mathbf{x}$, whereas QKV uses the normalized input $\tilde{\mathbf{x}}=\mathbf{x}/\rho(\mathbf{x})$. To share QKV's input, we rewrite the Gate/Up projection as
\begin{equation}
    \widetilde{\mathbf{W}}'
    \frac{\mathbf{x}'}{\rho(\mathbf{x}')}
    =
    \frac{\rho(\mathbf{x})}{\rho(\mathbf{x}')}
    \left[
        \underbrace{
            \widetilde{\mathbf{W}}'\tilde{\mathbf{x}}
        }_{\text{lookahead}}
        +
            \widetilde{\mathbf{W}}'
            \frac{\mathbf{u}}{\rho(\mathbf{x})}
    \right].
    \label{eq:system_simd_decomposition}
\end{equation}
% The lookahead and delta results are combined within DRAM, after which the host applies the factor $\rho(\mathbf{x})/\rho(\mathbf{x}')$.

As shown in Figure~\ref{fig:execution_computation_graphs}, execution begins with QKV and Gate/Up lookahead. After QKV completes, the CPU reads and reduces its partial sums, then executes attention and the Out projection to produce $\mathbf{u}$. Once both $\mathbf{u}$ and the lookahead result are available, Gate/Up delta accumulates into the lookahead result retained in DRAM. The combined result is then read out, reduced, and scaled by $\rho(\mathbf{x})/\rho(\mathbf{x}')$.

%%%%%%%%%%%%%
\subsubsection{MIMD Execution through Pipelining}
\label{sec:mimd_pipeline}

MIMD execution allows independently controlled resources to process different inputs with separate instruction streams. To exploit this parallelism, we apply lookahead--delta decomposition to both QKV and Gate/Up, overlapping each projection's delta with the successor's lookahead. Figure~\ref{fig:execution_computation_graphs} shows the dependencies enabling this execution. Each residual update makes inputs available for two independent computations. Once the attention update $\mathbf{u}$ is available, the GeMV components of Gate/Up delta and the next block's QKV lookahead are
\begin{equation}
    \underbrace{
        \widetilde{\mathbf{W}}_{G}^{(\ell)}\mathbf{u}
    }_{\text{Gate/Up delta}},
    \qquad
    \underbrace{
        \widetilde{\mathbf{W}}_{Q}^{(\ell+1)}
        (\mathbf{x}+\mathbf{u})
    }_{\text{next QKV lookahead}}.
    \label{eq:system_mimd_concurrency}
\end{equation}
%where the weight matrices incorporate the corresponding normalization gains; the final normalization factors are omitted here. 
% The second computation does not depend on the first computation's result, but the two consume different inputs, motivating execution on independent resources.

Figure~\ref{fig:execution_timeline} illustrates the resulting pipeline. QKV delta overlaps Gate/Up lookahead, while Gate/Up delta overlaps the next block's QKV lookahead. Each delta accumulates into its retained lookahead result and can begin once both its input update and that result are available. 

\begin{figure}[t]
    \centering
    \includegraphics[width=0.95\linewidth]
        {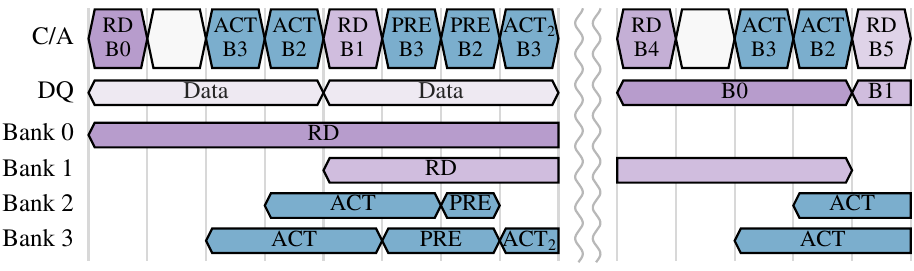}
    \caption{
        Overlapping data transfer and PUD computation.
    }
    \label{fig:compute_transfer_overlap}
\end{figure}

Lookahead can overlap with result readout because full DQ utilization does not necessarily imply full C/A utilization, as illustrated in Figure~\ref{fig:compute_transfer_overlap}. Each READ command initiates a data burst spanning multiple cycles, leaving C/A slots available between READ commands. The controller can use these slots to issue PUD commands to separate banks while DQ remains fully occupied with data transfers, subject to shared timing and resource constraints.

%%%%%%%%%%%%%%%%%%%%%%%%%%%
\subsection{Data Mapping and Tensor Partitioning}
\label{sec:partitioning_selection}

We model latency under each scheme's data mapping. Let $K$ denote the common input dimension of QKV and Gate/Up, and let $M_Q$ and $M_G$ denote their respective output dimensions. Let $H_Q$ and $H_G$ denote the CPU processing times following the QKV and Gate/Up, respectively. These include attention for $H_Q$, and Down for $H_G$.

\paragraph{\sysname{} (SIMD).}
Executing QKV and Gate/Up lookahead under a shared
instruction stream requires their weights to be
mapped within the same subarrays.
The paired projections use a common partition count $N$.
When packing does not lengthen the QKV instruction stream,
Gate/Up lookahead is fully hidden by QKV computation.
The decoder-block latency is then
\begin{equation}
\begin{aligned}
    T_{\mathrm{simd}}(N)
    ={}&
    \frac{\alpha K(\tau_Q+\tau_{\Delta,G})}{N} \\
    &+\beta(M_Q+M_G)N+H_Q+H_G,
\end{aligned}
    \label{eq:system_simd_latency}
\end{equation}
where $\tau_Q$ and $\tau_{\Delta,G}$ are the average
retained term counts per input element for direct QKV
and Gate/Up delta, respectively.
The optimum is
\(
    N_{\mathrm{simd}}^\star
    =
    \sqrt{
        \frac{\alpha K(\tau_Q+\tau_{\Delta,G})}
             {\beta(M_Q+M_G)}
    }.
\)

\paragraph{\sysname{} (MIMD).}
Executing QKV and Gate/Up with independent instruction streams requires their weights to be mapped to independently controlled resource groups, such as different banks. The two projections can use separate partition counts $N_Q$ and $N_G$. Lookahead overlaps the concurrent delta computation, reduction, and subsequent CPU processing. Let $L_G$ and $L_Q$ denote the execution times of Gate/Up lookahead and the next block's QKV lookahead, respectively. Assuming both computations in each pair can start concurrently, lookahead is fully hidden when
\begin{equation}
    L_j \le
    \frac{\alpha K\tau_{\Delta,i}}{N_i}
    +\beta M_iN_i+H_i,
    \label{eq:system_mimd_overlap}
\end{equation}
where $(i,j)\in\{(Q,G),(G,Q)\}$ and $\tau_{\Delta,i}$ is the average term count per input element for projection $i$'s delta. Under these conditions, the decoder-block latency is
\begin{equation}
    T_{\mathrm{mimd}}(N_Q,N_G)
    =
    \sum_{i\in\{Q,G\}}
    \left(
        \frac{\alpha K\tau_{\Delta,i}}{N_i}
        +\beta M_iN_i+H_i
    \right),
    \label{eq:system_mimd_latency}
\end{equation}
and the unconstrained continuous optimum for each projection is $N_{\mathrm{mimd},i}^\star =\sqrt{\alpha K\tau_{\Delta,i}/(\beta M_i)}$. Reducing the delta term count therefore favors fewer input partitions, also reducing partial-sum traffic.

\begin{figure}[t]
    \centering
    \includegraphics[width=0.95\linewidth]
        {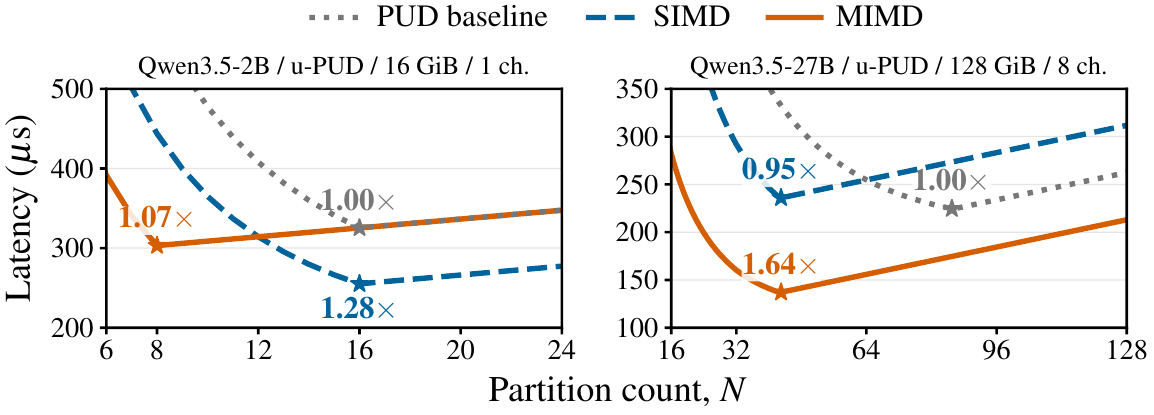}
    \caption{GeMV latency versus partition count.}
    \label{fig:scheme_partition_latency}
\end{figure}

\paragraph{Scheme Selection.} We first optimize input partitioning within each scheme, then select the scheme with the lowest predicted latency. The SIMD scheme is favorable when the paired projections fit within otherwise unused columns without requiring additional output tiles. The MIMD scheme can be favorable when sufficient independent resources and overlap time are available to hide lookahead. Figure~\ref{fig:scheme_partition_latency} shows how the latency-minimizing partition count and preferred scheme depend on the model--memory configuration.

%%%%%%%%%%%%%%%%%%%%%%%%%%%%%%%%%%%%%%%%%%%%%%%%%%%%%
\section{Evaluation}
\label{sec:evaluation}

We evaluate the decoding performance of \sysname{} and analyze how acceleration of the targeted projections contributes to the overall speedup. We also examine the utilization, quality--efficiency trade-off, and sensitivity to host-side optimizations and hardware configurations.

\begin{figure*}[t]
    \centering
    \includegraphics[width=0.95\linewidth]
        {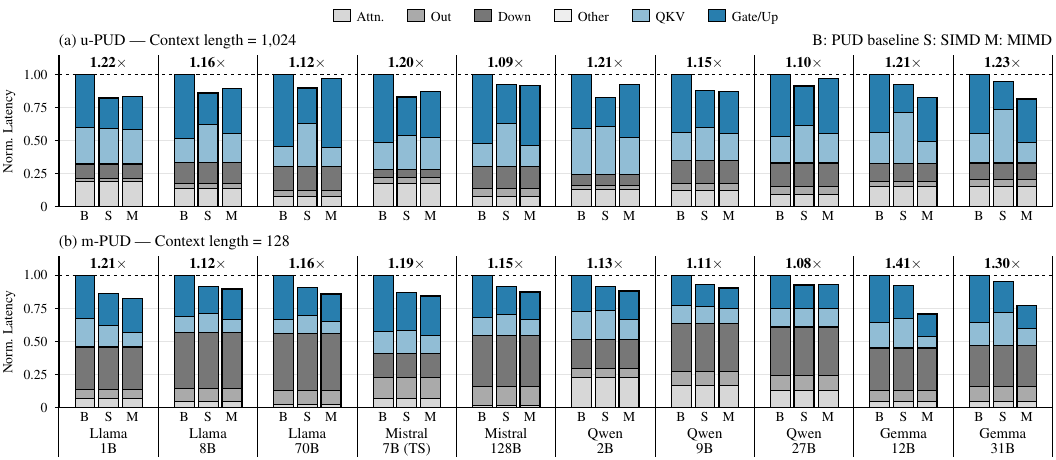}
    \caption{
        Latency breakdown, normalized to the PUD baseline. Stacked segments show operator contributions to the critical path, with host operators in gray and QKV and Gate/Up in blue. Thick outlines identify the fastest method in each group; annotations show its speedup over the corresponding PUD baseline. Lower bars indicate better performance.
    }
    \label{fig:eval_performance}
\end{figure*}

%%%%%%%%%%%%%%%%%%%%%%%%%%%
\subsection{Experimental Setup}
\label{ssec:setup}

\paragraph{Models and Hardware.}
We evaluate ten models from the Llama~\cite{grattafiori2024llama}, Mistral~\cite{mistral7b}, Qwen~\cite{team2026qwen3}, and Gemma~\cite{gemma4} families on both unmodified and modified DRAM. We adopt DDR4-2400 with 8\,GiB and 16 banks per rank, using two ranks per channel for all models. For each model, we select the smallest channel count from $\{1,2,4,8,16\}$ that permits all schemes to be placed under the u-PUD limit of 256 input elements per subarray. For host processing, we assume memory-bandwidth-bound execution with an effective bandwidth of 85\% of the peak memory bandwidth.

\paragraph{Quality Evaluation.} The term budget controls both approximation quality and PUD computation latency, linking our quality and performance evaluations. We evaluate its impact on model quality by applying the approximation to pretrained models and measuring perplexity on WikiText-2~\cite{merity2017pointer}. For each model and execution scheme, we select term budgets that satisfy the specified quality constraint for performance simulation.

\paragraph{Performance Evaluation.} We combine an analytical performance model with a PUD simulator built on Ramulator~2.0~\cite{luo2023ramulator}. We evaluate batch-one autoregressive decoding with a context length of 128 or 1024 tokens. The default weight precision is 4 bits; selected SIMD configurations use mixed precision with an average of 3--4 bits. Following prior architectural evaluations~\cite{park2024attacc,heo2024neupims}, we report decoding latencies per generated token across all Transformer blocks, excluding prefill, embedding lookup, final normalization, the language-model head, and sampling. We assume input sparsification for the host-side Out and Down projections, allowing weight transfers and computations associated with pruned input elements to be skipped~\cite{liu2025training}. Guided by the sparsity levels reported in that work, we assume input sparsities (i.e., fractions of input elements pruned) of 70\% for Out and 65\% for Down across all models, with Down-input sparsity of 90\% for Mistral~7B (TS)~\cite{song2024turbo}.

%%%%%%%%%%%%%%%%%%%%%%%%%%%
\subsection{Decoding Performance}
\label{ssec:decoding_performance}

For each model and execution scheme, we select term budgets that limit the perplexity increase to at most 5\% relative to the quantized model without term truncation.
Figure~\ref{fig:eval_performance} shows decoding latency and its breakdown, normalized to the PUD baseline. Taking the faster of SIMD and MIMD for each model, \sysname{} achieves geometric-mean speedups of $1.17\times$ on u-PUD and $1.18\times$ on m-PUD. The maximum speedups are $1.23\times$ for Gemma~31B on u-PUD and $1.41\times$ for Gemma~12B on m-PUD. 

\begin{figure}[t]
    \centering
    \includegraphics[width=0.95\linewidth]{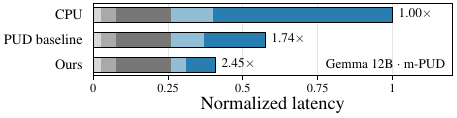}
    \caption{Decoding latency normalized to CPU.}
    \label{fig:eval_cpu_comparison}
\end{figure}

\begin{figure*}[t]
    \centering
    \includegraphics[width=0.95\linewidth]{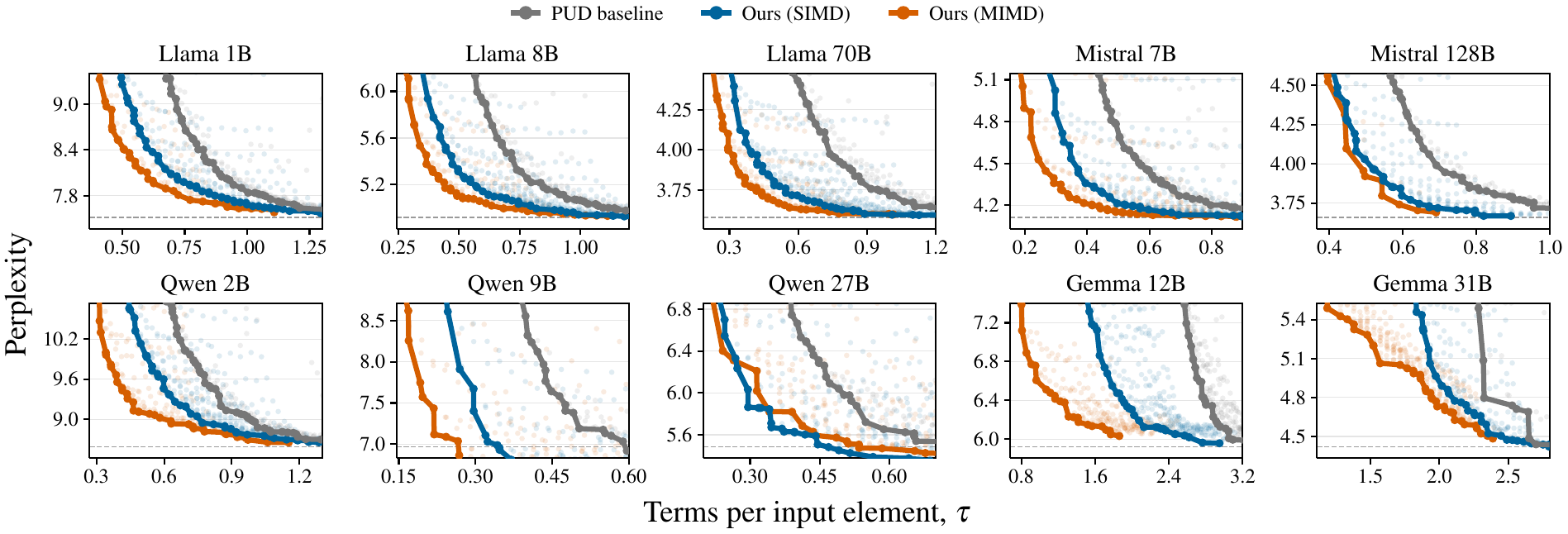}
    \caption{Quality--efficiency trade-offs: perplexity on WikiText-2 versus average terms per input element on the critical path. \sysname{} achieves comparable perplexity
    with fewer terms. Faint points show evaluated settings; highlighted points trace the Pareto frontiers.}
    \label{fig:eval_quality_efficiency}
\end{figure*}

\begin{figure}[t]
    \centering
    \includegraphics[width=0.95\linewidth]{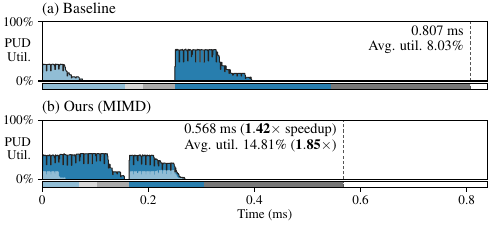}
    \caption{Lookahead overlaps projection computation, increasing mean utilization while reducing block latency.} % on m-PUD. Lookahead overlaps projection computation, increasing mean utilization while reducing block latency.}PUD utilization during Gemma~12B decoder block. 
    \label{fig:eval_parallelism_utilization}
\end{figure}
The latency breakdown illustrates both the benefit of projection acceleration and the limit imposed by host processing. For Gemma~12B on m-PUD, the QKV and Gate/Up projections account for 54.6\% of the PUD baseline latency, while the remaining operations account for 45.4\%. MIMD accelerates these projections by $2.13\times$. With the remaining latency unchanged, this yields an overall speedup of $1/(0.454+0.546/2.13)\approx1.41\times$ over the PUD baseline, corresponding to a $2.45\times$ speedup over modeled CPU execution (Figure~\ref{fig:eval_cpu_comparison}).

Figure~\ref{fig:eval_parallelism_utilization} illustrates PUD parallel resource utilization during one Gemma~12B decoder block on m-PUD. Utilization is defined as the fraction of available column resources occupied by PUD computation, including lookahead, estimated from Ramulator command traces and mapped column counts and averaged over $1\,\mu$s intervals. %Colors follow the operator categories in Figure~\ref{fig:eval_performance}; the strips below the curves show foreground execution, and dashed lines mark block completion. %Lookahead overlaps Gate/Up computation with QKV execution and precomputes the next block's QKV during Gate/Up execution. 
For this block, mean utilization rises from $8.03\%$ to $14.81\%$ ($1.85\times$), while latency decreases from $0.807$ to $0.568$\,ms ($1.42\times$ speedup).

%%%%%%%%%%%%%%%%%%%%%%%%%%%
\subsection{Quality--Efficiency Trade-off}
\label{ssec:quality_efficiency}

The preceding performance comparison uses a fixed quality constraint. The analysis in Section~\ref{sec:delta_approximation} suggests that \sysname{} requires fewer terms than direct computation at the same input-approximation error. Assuming comparable input-approximation errors have similar effects on model quality, we expect this term reduction to also hold at comparable perplexity.

Figure~\ref{fig:eval_quality_efficiency} shows the results. The horizontal axis measures the average number of nonzero terms per input element on the critical path, assuming fully hidden lookahead. %This includes direct QKV and Gate/Up for the PUD baseline, direct QKV and Gate/Up delta for SIMD, and QKV and Gate/Up delta for MIMD. 
Across all evaluated models, the proposed schemes achieve comparable perplexity with fewer terms than the PUD baseline, supporting our analysis. Since projection computation latency scales with the number of nonzero terms on the critical path, these results indicate a better quality--efficiency trade-off.
The extent of the reduction varies by model: MIMD requires substantially fewer terms than SIMD at comparable perplexity on Gemma~12B, whereas SIMD offers a comparable trade-off to MIMD on Qwen~27B and Mistral~128B.

%%%%%%%%%%%%%%%%%%%%%%%%%%%
\subsection{Sensitivity Analysis}
\label{ssec:sensitivity}

We vary memory configuration, context length, and Down-input sparsity to examine how PUD resources and CPU processing costs affect the benefits of \sysname{}.

\begin{figure}[t]
    \centering
    \includegraphics[width=0.95\linewidth]{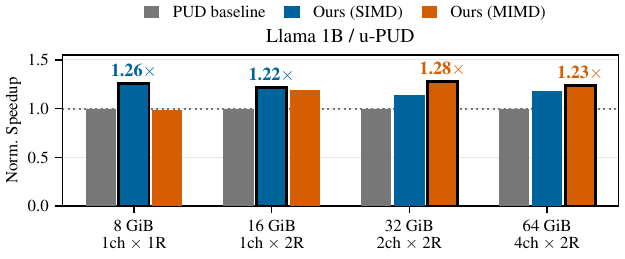}
    \caption{ Increasing memory parallelism can shift the preferred execution scheme from SIMD to MIMD for the same model. }
    \label{fig:eval_memory_configuration}
\end{figure}
\paragraph{Memory Configuration.} The preferred execution scheme depends on model dimensions and available memory parallelism. Figure~\ref{fig:eval_memory_configuration} examines this dependence by varying channel and rank counts for Llama~1B on u-PUD. SIMD is preferable in the 8 and 16\,GiB configurations, exploiting unused columns without requiring separate banks for lookahead. Additional independent resources favor MIMD in the 32 and 64\,GiB configurations by reducing competition between lookahead and delta computation. For example, SIMD achieves a $1.26\times$ speedup over the PUD baseline at 8\,GiB, while MIMD achieves $1.28\times$ at 32\,GiB.

\begin{figure}[t]
    \centering
    \includegraphics[width=0.95\linewidth]{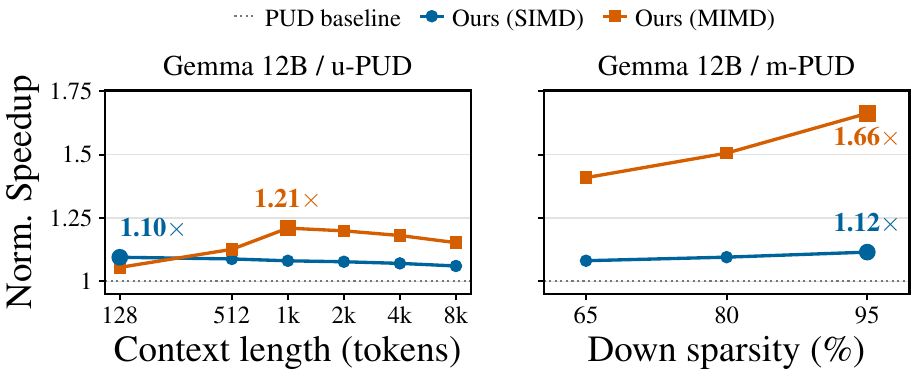}
    \caption{Longer CPU processing limits overall speedup, while shorter CPU processing can leave lookahead exposed.}
    \label{fig:eval_host_sensitivity}
\end{figure}
\paragraph{CPU Processing.}
We examine how CPU processing time affects decoding speedup by varying context length on u-PUD and Down-input sparsity on m-PUD for Gemma~12B (Figure~\ref{fig:eval_host_sensitivity}).
At short contexts, the small attention cost allows SIMD to achieve its highest observed speedup of $1.10\times$ at 128 tokens. MIMD, however, has insufficient time to hide Gate/Up lookahead. Longer contexts provide more overlap time, allowing MIMD to reach its highest observed speedup of $1.21\times$ at 1k tokens. Beyond this point, attention computation and KV-cache accesses account for an increasing share of decoding latency, reducing overall speedup despite continued projection acceleration. Our approach is therefore complementary to techniques that accelerate attention~\cite{tang2024quest,liu2024kivi}.
The Down-input sparsity sweep models potential CPU-side acceleration through activation sparsification~\cite{song2024turbo}. At the highest evaluated sparsity of 95\%, both schemes achieve their largest speedups: $1.12\times$ for SIMD and $1.66\times$ for MIMD. Although Down execution becomes shorter, QKV lookahead remains hidden. % by Gate/Up delta computation, allowing MIMD to benefit from the reduced CPU cost.

%%%%%%%%%%%%%%%%%%%%%%%%%%%%%%%%%%%%%%%%%%%%%%%%%%%%%
\section{Related Work}

\paragraph{In-Memory Computing Systems for LLM Inference.} PIM systems improve execution efficiency through data placement and address mapping~\cite{ibrahim2024pimnast,seo2025facil}, as well as operator assignment, partitioning, and execution overlap~\cite{zhou2023overlapim,heo2024neupims,gu2025cent,he2025papi}. Hardware/software support for arithmetic synthesis and fine-grained resource control enables flexible in-DRAM execution~\cite{hajinazar2021simdram,oliveira2024mimdram}. Further optimizations adapt precision, data representations, and arithmetic implementations to reduce computation latency~\cite{oliveira2025proteus}, eliminate input preparation and output bit-transposition overheads~\cite{kubo2025mvdram}, and remove runtime data transposition while enabling bit-level pipelining~\cite{oliveira2026pipedram}. In contrast, \sysname{} restructures the computation itself by exploiting LLM-specific properties.

\paragraph{Parallelizing Dependent Computations.} Prior work exposes parallelism in dependent computations through finer task granularity, pipelining, and computational reformulation. Task-based execution advances computations as their inputs become available, while software pipelining overlaps stages without violating data dependences~\cite{lam1988software,ottoni2005automatic,buttari2009class}. Other approaches use algebraic transformations~\cite{blelloch1990prefix}, predict unavailable values and subsequently validate them~\cite{lipasti1996exceeding}, or relax selected dependences under application-specific accuracy constraints~\cite{udupa2011alter,misailovic2013parallelizing}. For neural networks, prior work either evaluates provisional intermediate states in parallel and refines them iteratively~\cite{song2021accelerating,danieli2023deeppcr} or exploits similarity between layer inputs~\cite{zou2024cqil}. At the token level, candidate generation and parallel verification reduce sequential decoding steps~\cite{leviathan2023fast,fu2024break}. 
Our work also algebraically reformulates dependent computations using redundancy in LLMs.

\paragraph{Redundancy-Aware Approximate Inference.} Redundancy and error tolerance in neural networks allow selected computations to be skipped or performed at lower precision with limited impact on model quality. Mixed-precision quantization allocates precision according to its impact on model quality~\cite{yao2021hawqv3,he2023ptqd,zheng2026mixllm}, while precision-adaptive execution translates reduced precision into lower execution cost~\cite{judd2016stripes,sharma2018bitfusion, sabri2024redy}. Redundancy across LLM layers~\cite{lad2024remarkable,gromov2024unreasonable} motivates layer omission or consolidation~\cite{delcorro2023skipdecode,song2024sleb,yang2024laco,men2024shortgpt}, approximation or reuse of intermediate representations~\cite{din2023jump, razzhigaev2024secretly,chen2025streamline,mu2024crosslayer}, and input substitution for concurrent layers~\cite{zou2024cqil}. Spatial and temporal redundancy also enables differential computation~\cite{mahmoud2018diffy,kim2025ditto}. Our work exploits inter-layer redundancy to approximate the corrections, rather than skipping layers, without model retraining or architectural modifications.

%%%%%%%%%%%%%%%%%%%%%%%%%%%%%%%%%%%%%%%%%%%%%%%%%%%%%
\section{Conclusion}
\label{sec:conclusion}

We presented \sysname{}, an algorithm--system co-design whose algorithm exploits LLM layer redundancy to expose inter-layer parallel work for efficient execution on PUD. Its system design improves SIMD and MIMD utilization through packing and pipelining, respectively.
\sysname{} improves the quality--efficiency Pareto frontier. Performance simulations show up to $1.41\times$ end-to-end decoding speedup over the PUD baseline. These results demonstrate the importance of restructuring computation around model dependencies and numerical properties to use PUD resources effectively, complementing optimizations of individual operations.
Overall speedup remains constrained by operations executed on the CPU. Coordinating their acceleration with PUD execution offers opportunities for further improvement. Future work will explore other PIM architectures and investigate opportunities and conditions for applying the approach to model structures such as MoE.

\bibliography{main}
\bibliographystyle{mlsys2025}

%%%%%%%%%%%%%%%%%%%%%%%%%%%%%%%%%%%%%%%%%%%%%%%%%%%%%%%%%%%%%%%%%%%%%%%%%%%%%%%
%%%%%%%%%%%%%%%%%%%%%%%%%%%%%%%%%%%%%%%%%%%%%%%%%%%%%%%%%%%%%%%%%%%%%%%%%%%%%%%
% SUPPLEMENTAL CONTENT AS APPENDIX AFTER REFERENCES
%%%%%%%%%%%%%%%%%%%%%%%%%%%%%%%%%%%%%%%%%%%%%%%%%%%%%%%%%%%%%%%%%%%%%%%%%%%%%%%
%%%%%%%%%%%%%%%%%%%%%%%%%%%%%%%%%%%%%%%%%%%%%%%%%%%%%%%%%%%%%%%%%%%%%%%%%%%%%%%
\clearpage
\appendix

%%%%%%%%%%%%%%%%%%%%%%%%%%%%%%%%%%%%%%%%%%%%%%%%%%%%%
\section{PUD Execution Model}
\label{app:pud_execution}

This appendix describes the m-PUD operation costs and their relationship to $\alpha$, $\beta$, and $P$ in our latency model. We adopt the addition sequence from SIMDRAM~\cite{hajinazar2021simdram} and the primitive-operation timings from the MIMDRAM implementation~\cite{oliveira2024mimdram}. % \footnote{ \href{https://github.com/CMU-SAFARI/MIMDRAM/tree/23495f10950d891a95a0b8a05d0a6a88e92de154/gem5/src/mem} {MIMDRAM repository, commit \texttt{23495f1}}. }

\paragraph{Primitive Operations and Timing.}
Modified DRAM supports ACT--PRE (AP) and
ACT--ACT--PRE (AAP) sequences.
AP activates the addressed row or row group and
then precharges the bank.
AAP additionally copies the activation result
to a destination row or row group before precharging.
Selecting designated compute rows enables these
sequences to perform majority operations and preserve
their results~\cite{seshadri2017ambit}.

The public DDR4-2400 configuration specifies
$t_{\mathrm{CK}}=0.833$\,ns,
$t_{\mathrm{RAS}}=32$\,ns, and
$t_{\mathrm{RP}}=14.16$\,ns.
Overlapping the two activations in AAP adds
$t_{\mathrm{ov}}=4$\,ns.
The resulting primitive latencies are
\begin{align}
    T_{\mathrm{AP}}
    &= t_{\mathrm{RAS}}+t_{\mathrm{RP}}
     = 46.16\,\mathrm{ns}, \\
    T_{\mathrm{AAP}}
    &= t_{\mathrm{RAS}}+t_{\mathrm{ov}}
       +t_{\mathrm{RP}}
     = 50.16\,\mathrm{ns}.
\end{align}
These isolated latencies include final precharge
recovery and exclude queuing and resource contention.

\paragraph{Addition and Per-Term Computation Cost.}
The addition loop executes six AAPs and two APs per bit, including operand preparation and result preservation~\cite{hajinazar2021simdram}. Thus, one full-adder step takes
\begin{equation}
    T_{\mathrm{FA}}
    =6T_{\mathrm{AAP}}+2T_{\mathrm{AP}}
    =393.28\,\mathrm{ns}.
\end{equation}
Including one initialization AAP, the DRAM-operation time for a $b$-bit addition is
\begin{equation}
    T_{\mathrm{add}}(b)
    =T_{\mathrm{AAP}}+bT_{\mathrm{FA}}
    =(50.16+393.28b)\,\mathrm{ns}.
\end{equation}
% The coefficient $\alpha$ denotes the average cost of accumulating one retained input term, accounting for the number of bit positions processed.

\paragraph{Reduction Cost.} The coefficient $\beta$ captures partial-sum readout and CPU aggregation per output element per input partition. With weight precision $b_w$ and bit-plane partial-sum precision $b_{\mathrm{ps}}$, readout transfers $b_wb_{\mathrm{ps}}/8$ bytes per output element per partition. At effective readout bandwidth $\mathrm{BW}$, this contributes $b_wb_{\mathrm{ps}}/(8\mathrm{BW})$ to $\beta$; the remaining contribution accounts for CPU aggregation.

\paragraph{Effective MIMD Parallelism.}
Computations in different banks can overlap, but banks within a channel share the C/A bus. In our analytical model, each ACT or PRE occupies one C/A cycle. For a repeated operation with isolated duration $L$ and $C$ commands, the command-bandwidth bound on concurrency at the isolated execution rate is
\begin{equation}
    P
    =\frac{L}{C\,t_{\mathrm{CK}}}.
\end{equation}
Expanding each AAP into three commands and each AP into two, the full-adder loop requires $C=6\times3+2\times2=22$ commands. Its per-channel bound is therefore
\begin{equation}
    P
    =\frac{393.28}{22\times0.833}
    \approx21.46.
\end{equation}
% This value is an average command-demand bound for the repeated full-adder loop, rather than a physical bank count or guaranteed achievable concurrency.

%%%%%%%%%%%%%%%%%%%%%%%%%%%%%%%%%%%%%%%%%%%%%%%%%%%%%%%%%%%%%%%%%%%%%%%%%%%%%%%
%%%%%%%%%%%%%%%%%%%%%%%%%%%%%%%%%%%%%%%%%%%%%%%%%%%%%%%%%%%%%%%%%%%%%%%%%%%%%%%

\end{document}